# Wolf in Sheep's Clothing – The Downscaling Attack Against Deep Learning Applications

Qixue Xiao[1], Kang Li[2], Deyue Zhang[1], Yier Jin[3]
[1] Qihoo 360 Security Research Lab
[2]University of Georgia
[3]University of Florida

***Abstract*—This paper considers security risks buried in the data processing pipeline in common deep learning applications. Deep learning models usually assume a fixed scale for their training and input data. To allow deep learning applications to handle a wide range of input data, popular frameworks, such as Caffe, TensorFlow, and Torch, all provide data scaling functions to resize input to the dimensions used by deep learning models. Image scaling algorithms are intended to preserve the visual features of an image after scaling. However, common image scaling algorithms are not designed to handle human crafted images. Attackers can make the scaling outputs look dramatically different from the corresponding input images.**

**This paper presents a downscaling attack that targets the data scaling process in deep learning applications. By carefully crafting input data that mismatches with the dimension used by deep learning models, attackers can create deceiving effects. A deep learning application effectively consumes data that are not the same as those presented to users. The visual inconsistency enables practical evasion and data poisoning attacks to deep learning applications. This paper presents proof-of-concept attack samples to popular deep-learning-based image classification applications. To address the downscaling attacks, the paper also suggests multiple potential mitigation strategies.**

## I. Introduction

Deep learning applications are continuously gaining popularity in both academic and industry in recent years. Advances in GPUs and deep learning algorithms along with large datasets allow deep learning algorithms to address real-world problems in many areas, from image classification, health care prediction, auto game playing, to reverse engineering.

Along with this increasing popularity comes with new security concerns to deep learning applications. Data poisoning attacks [1, 4], such as mixing mislabeled data or polluted samples in training datasets, have drawn substantive attention. Researchers [10, 16] have shown that deep learning applications, such as imaging recognition and self-driving, can be manipulated to make wrong decisions on benign inputs due to data poisoning.

In practice, data poisoning attacks are not the main concern to deep learning application developers and users. One possible reason is that users often believe the risk can be largely mitigated by having human carefully inspect the training data. Using image classification applications as examples, people generally believe they can visually detect those mislabelled images (e.g. having a human verify an image label by looking at it).

This paper presents *downscaling attacks*, which target a security risk buried in the data processing pipeline of deep learning applications. Downscaling attacks take advantage of a limitation in the data flow processing in deep learning applications – deep learning model are typically trained on well formatted data. For design simplicity, most deep learning applications use fixed size input scales. The designer of deep neural networks assume both training data and application inputs fall on the same scale. In practice, many deep learning applications often need to deal with inputs that are formatted in different scales. Arbitrary inputs therefore need to be resized to the model scale in order to be used by the deep learning neural network. To support this particular need, popular deep learning frameworks all offer rescaling functions; and for most deep learning applications we examined, data rescaling is a common component in their data processing pipeline.

In a downscaling attack, attackers feed input images containing camouflage data to a deep learning application. The camouflage data are arranged in an input image such that they will be filtered or changed by the scaling algorithm. The result is an image with a different visual presentation being fed to the application's deep learning neural network. Consequently this attack creates visually deceiving effect and potentially allows attackers to launch more practical data poisoning and evasion attacks [1, 4, 11, 13, 20].

In the paper, we investigate the common scaling implementations in three popular deep learning frameworks: Caffe [9], TensorFlow [7], and Torch [15]. We find that all the default data scaling algorithms are vulnerable to the downscaling attack. Attackers can injects poisoning or deceiving data to training samples or input data. The injected camouflaged data are visible to users but are discarded by scaling functions, and thus these data are eventually not used by deep learning algorithms. For example, a human considers an image (with camouflage) only contains sheep, but the image becomes a picture of wolf after rescaling. For deep learning applications that take the camouflaged sheep image, it would be the wolf image eventually being proceeded by applications.

This paper describes our preliminary study on the threat related to data scaling, and we provide proof of concept inputs to illustrate the potential danger of downscaling attacks. Through our preliminary study of the commonly used scaling options in deep learning applications, we make the following contributions:

- This paper presents a preliminary study of the data scaling options in popular deep learning frameworks.
- We find that data scaling is actually a common need in deep learning applications.
- We show the risk of downscaling attacks is real, and we provide multiple proof-of-concept images targeting

popular deep learning applications.
- To reduce the potential threats from downscaling attacks, we offer a few mitigation strategies, such as input filtering, robust scaling implementations, and downscaling attack detections.

## II. Downscaling Attack Examples

To illustrate the potential risks and the deceiving effect of downscaling attacks, we present a few sample attack inputs in this section.

Figure 1 presents the first group of examples that were crafted for image classification applications. The targeted application used here is the *cppclassification*[8] sample released along with the Caffe framework.

To show the effect of these images on deep learning applications, we used the GoogleNet model provided by the BAIR lab of BVLC [6], which assumes the input data is in the scale of 224*224. When an image with a different size is provided, the application uses the *resize()* function in the Caffe framework to rescale the input to the size used by the model (224*224). The exact classification setup details and the program output are presented in the appendix of this paper.

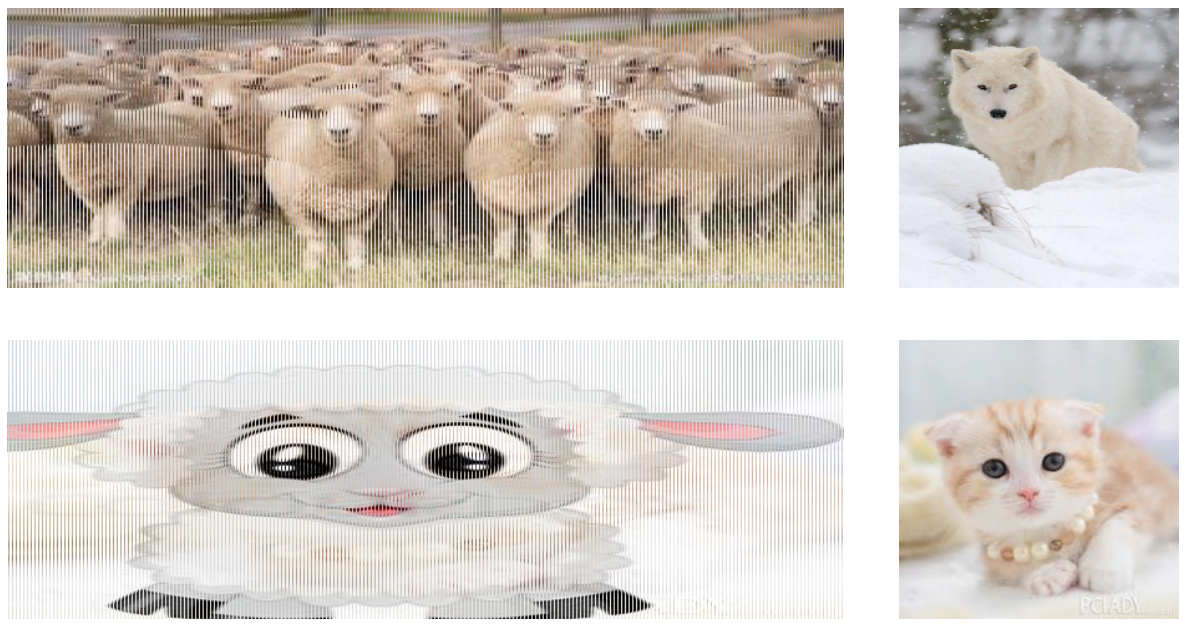

Fig. 1: Two Examples Showing Sample Effect of *Downscaleing* Attack. (Left-side: What Human See; Right-side: What ML Models See)

We specially crafted input images that are in a different size (224*672). The images on the left column of Figure 1 are inputs to deep learning applications. The images on the right column are the output of the scaling function, and thus are the *effective images* used by the deep learning algorithms.

While the input on the left column are sheep, the deep learning model takes the images on the right column as the real inputs. The corresponding image are classified into instances of “White Wolf” and “lynx Cat” respectively.

TABLE I: PoC Sample Image Information

| Image File | MD5 Checksum | Image Size |
|---|---|---|
| wolf-in-sheep.png | fce6aa3c1b03ca4ae3289f5be22e7e65 | 224*672 |
| wolf.png | ed2d62fbb8720d802c250fb386c254f6 | 224*224 |
| cat-in-sheep.png | 6d489fe75f74ad32e8ada01ff7da9450 | 224*672 |
| cat.png | 01820651a302fb3730ac0a5ffd95c23c | 224*224 |

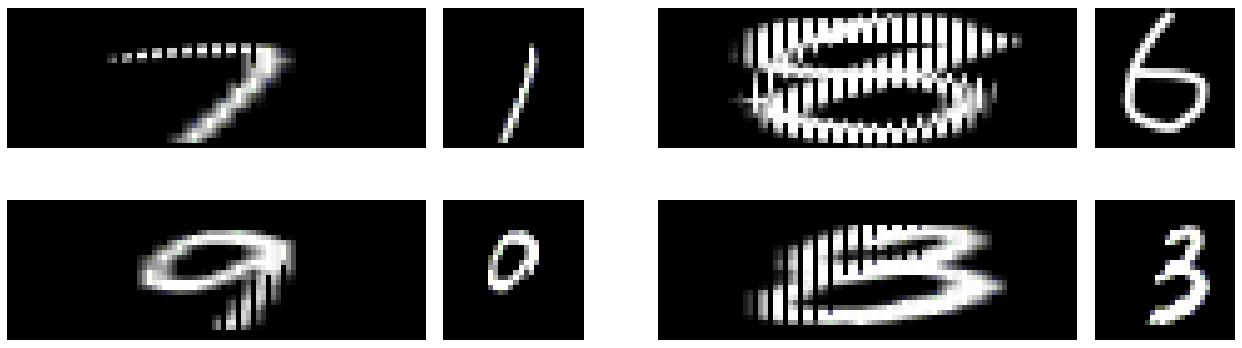

Fig. 2: Four pairs of *Downscaling* attack examples for the MNIST application (Within each pair, left-side: What Human See; right-side: What ML Models See)

Figure 2 shows the attack examples using the MNIST handwriting digits recognition [21] as the target application. Four pairs of images are presented in this figure. Within each pair, the crafted inputs (visible to human) are in the left column and the scaling results (visible to models) are in the right column. The figure shows that the images of digits change dramatically after the scaling call.

## III. Data Scaling and Deep Learning Applications

Data scaling is not limited to images, and various resampling approaches can be applied to voice and other types of data. For simplicity, this paper focuses on the scaling effort for image data. Image scaling refers to the resizing of a digital image [19]. When scaling an image, the downscaling (or upscaling) process generates a new image with a lower (or higher) number of pixels compared to the original image. In the case of decreasing the pixel number (downscaling), this usually results in a visible quality loss.

### A. The Practical Need of Rescaling

Although scaling is not a necessary component for deep learning algorithms, data scaling is actually quite common in deep learning applications. First, open-input applications, such as image classifications as an Internet service, would have to involve scaling in their data processing pipeline. For design simplicity and manageable training process, a deep learning neural network model usually handles a fixed scale input size. Although a deep learning neural network can be designed with layers that take the largest possible scale among all possible inputs, this approach is inefficient in terms both training and classification.

For those deep learning applications that take input data from a fixed input source, such as sensors like video camera, the input data format are naturally uniform. Even in such situation, resize is used in certain cases.

One common situation we observe is the use of pretrained model. For example, NVIDIA offers multiple self-driving sample models [5], and all these models use an input size 200x66. However, on the recommended camera [12] specification provided by NVIDIA, the image generated are mostly of the size 1928x1208. Therefore, for system developers that do not want to design and train their own models, they are likely have to use scaling functions somewhere in their data processing pipeline. Recent research work, such as the sample applications used in the study of DeepXplore [14], also show that the resizing operation is commonly used in self-driving applications to fit camera output size to the size of models.

### B. Data Scaling in Deep Learning Image Applications

Most deep learning frameworks provide data scaling functions. We examine the sample programs come with popular deep learning frameworks, such as Tensorflow, Caffe, and Torch, and we found many programs call scaling functions in their data processing pipelines. For example, we found that the image classification application *cppclassification* [8] in Caffe framework uses data scale in function *Classifier::Preprocess*. Both C++ and Python image recognition demo using Tensorflow use function *ResizeBilinear* for data scaling[17]. Torch7 and PyTorch are also using data scaling in their tutorials [2, 18]. The corresponding code snippets are provided in the Appendix.

Besides examples from popular deep learning frameworks, we also found popular deep learning tools, such as deepdetect[3], use data scaling in their data processing pipeline.

TABLE II: Data scale algorithm in deep learning frameworks

| DL Framework | Library | Default Scaling Algorithm |
|---|---|---|
| Caffe | opencv | Bilinear |
| Tensorflow | python-opencv | Bilinear |
| Tensorflow | pillow | Neareast |
| Tensorflow | tf.image | Bilinear |
| Torch | torch-opencv | Bilinear |
| Torch | lua.image | Bilinear |

### C. Image Scaling Algorithms in Deep Learning Frameworks

Image scaling is a well studied field, and there exist multiple approaches when scaling an image from one size to another. The most common used algorithms include Nearest-Neighbors, Bilinear, and Bicubic interpolation [19].

The nearest-neighbor interpolation algorithm uses a relatively simple approach. It uses the value of the nearest point in the scaling process. To take neighboring points into consideration, linear interpolation approach uses a mathematical mean to represent a region. Image is often considered as 2D data, and bilinear approach is an extension of linear interpolation on a 2D grid.

Most of deep learning frameworks use bilinear interpolation as the default data scale algorithm. We inspected the implementation of scaling functions in deep learning frameworks. Table II shows the data scale algorithms used by Caffe, TensorFlow and Torch. Most frameworks support Nearest-Neighbors, Bilinear, Bicubic algorithms along with a few others depends on the specific framework.

## IV. The Downscaling Attack

This section provides an overview of the downscaling attack method against deep learning image applications.

### A. The Attack Goal and Constraints

The intent of image scaling algorithms is to adjust the image size while preserve visual features after scaling. The purpose of downscaling attack is to create effect against this intent. A downscaling attack is to craft images that produce downscaled outputs with dramatic visual differences from the inputs.

Although scaling in either direction (up and down scaling) could result in distortions and potentially deceiving effect, we only discuss the downscaling functions in this paper.

We can consider the scaling effect as a function $F$, which convert a source image in a $m*n$ dimension ($in[m][n]$) to a new image in different scale ($out[m'][n']$).

$$F(in[m][n]) = out[m'][n'](m' <= m, n' <= n) \quad (1)$$

Although not all downscaling attacks need to generate a specific image, for simplicity of discussion, we consider the following attack goal – Given a specific image target, a successful downscaling attack need to meet the following three constraints:

1) The attacker needs to compose an image ($crftImg[m][n]$) so that the output of $F(crftImg[m][n])$ is the target Image ($dstImg[m'][n']$);
2) The crafted image $crftImg[m][n]$ needs to be visually meaningful to human,
3) The crafted image $crftImg[m][n]$ and target image $dstImg[m'][n']$ contain different semantical meanings.

### B. The Nearest-neighbor Algorithm and Camouflage Regions

Downscaling algorithms generally do not take specially crafted "malicious" inputs into consideration when considering how to extract "sampling" data from input images. Here we take the simplest scaling algorithm as an example to demonstrate the downscaling attack.

Figure 3 illustrates the process of converting a 4x4 input image to the scale of 2x2. In this scaling process, only one fourth of the input image needs to contain the exact value of the output image pixels. This scaling process leaves three quarters of input data under attacker's control. We call these areas *camouflaging regions*. Regardless what values were stored in these camouflaging regions, the scaling output is the same. Therefore, all the attackers need to do is to use these camouflaging regions to create deceiving effect.

Certainly what exact content to put in the camouflage areas depends on the target image and the deceiving effect the attacker wants to achieve. Currently we are working on software tools to automatic interleaving pre-captured camouflage images over a given target image.

### C. Parameters and Consideration in Downscaling Attacks

To use downscaling attacks to achieve deceiving effects, attackers have to consider multiple factors, such as what to put in the camouflage regions and what size to pick for the input images.

First, although the early examples of nearest-neighbors algorithm leave camouflaging regions for attacker to inject arbitrary data, other scaling algorithms might impose different constraints. For example, Bilinear scaling algorithms requires a region's weighted average pixel values to be equivalent to the value of a corresponding pixel in the target image. Nevertheless, even with these constraints, attackers are given opportunities to provide content that will be partially discarded by the scaling function. In fact, the sample attack images in the previous sections are crafted against Bilinear scaling algorithms, and they can successfully trigger misclassifications in deep learning applications.

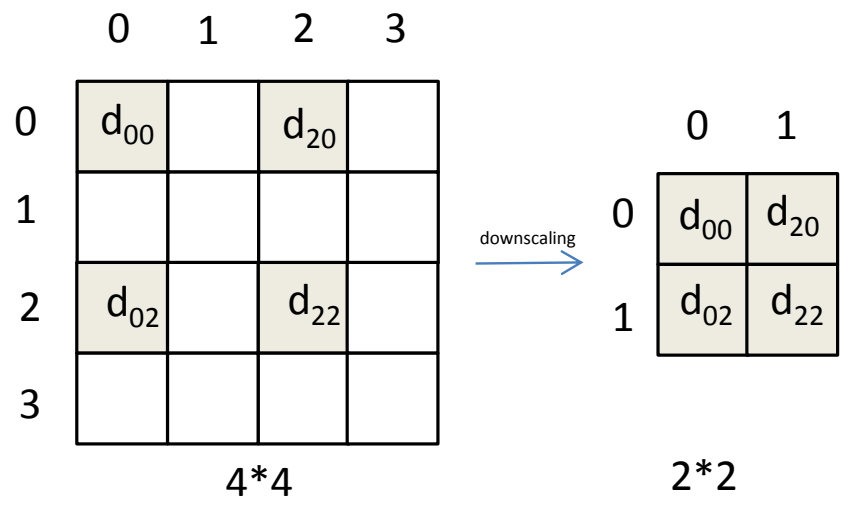

Fig. 3: Nearest-neighbor interpolation Illustration.

Second, the amount of data controlled by attackers is proportionally related to the scaling factor. An image that is much larger than the scale used by deep learning models generally provides more space for attackers to create deceiving effect. However, images with extremely uncommon sizes are likely trigger warnings to users who inspect inputs. Therefore, attackers need to consider the trade-off between image sizes and the levels of awkwardness potentially raised by the abnormal sizes.

## V. Discussion

### A. Practical Challenges for Applying Downscaling Attacks to Sensor-based Applications

Downscaling attacker certainly only affects applications that use scaling in the data processing pipeline. It does not threat applications that only consume data in the same scale as used by the deep learning models.

Certainly, even resize is used in deep learning applications, attackers still need to find out ways to inject crafted inputs in order to generate downscaling attacks. Launching such attack directly from the analog domain is challenging but not impossible. Moreover, sensors themselves potentially can be potentially compromised and malicious sensor inputs can be generated in those cases to attack the backend deep learning implementations. In the case of a compromised sensor, certainly it can directly feed more damaging input to the backend systems. However, downscaling attacks would still be preferred because they provide an additional level of deception.

### B. Mitigation of Downscaling Attacks

Deep learning application designers have multiple options to mitigate the risks from downscaling attacks. We have considered the following possible approaches: 1) limiting the input size, 2) using robust scaling algorithms, and 3) detecting dramatic changes from scaling,

The most straightforward way to avoid downscaling attack is to ignore inputs that do not match the scale used by the deep learning models. This approach fits well to applications that deal with input generated by sensor or other applications control by users. However, as we discussed above, not all applications can take this solution, and many deep learning applications often have to deal with input generated (and thus controlled) by Internet users.

Another possible solution is to adopt a robust scaling option. Although we have shown that the default scaling implementation in deep learning frameworks are vulnerable to downscaling attacks, there are scaling algorithms (such as Bicubic interpolation) that are more robust in terms of preserving the content of the original images, and thus are less vulnerable to attacks. Among the major frameworks we have inspected, Caffe, TensorFlow, and Torch all allow applications to choose different scaling algorithms through a parameter of the resize function.

The third viable solution is to detect dramatic changes in input features during the scaling process. The downscaling attack's deceiving effect is caused by a dramatic change to the image content before and after the scaling action. Therefore, applications can potentially detect such attacks by measuring features that are related to visual presentation. For example, the color histogram is likely changed in the case that the rescaled image is dramatically different from the original image.

## VI. Conclusion

The purpose of this work is to raise awareness of the security threats buried in the data processing pipeline in deep learning applications.

Deep learning applications usually make strong assumptions of a uniformed input data scale. All the training and classification algorithms focus on processing well formatted data. Although there are situations where data inputs uniformly match the scale used by the background models, many deep learning applications (such as image classifications) are unavoidable to process data that are in different scales.

Most deep learning frameworks provide scaling and resizing utilities to convert data to the scale used by deep learning models. This data processing stage is vulnerable to the downscaling attack presented in this paper. We showed that attackers can prepare visually deceiving images to deep learning applications. The result is a mismatch between what users see and what deep learning applications use. We studied multiple deep learning frameworks and showed that most of their default scaling functions are vulnerable to such attack. We hope our preliminary results in this paper can remind researchers to not ignore threats in the data processing pipeline and actively look for ways to detect risks in deep learning applications.

## Appendix

This appendix provides code snippet of deep learning frameworks using data scaling, and the brief information related to our experimental setup. All the software implementation and deep learning model were obtained online without change.

### A. *Code snippet of data-scaling used by deep learning frameworks and applications (The use of resizing or scaling function are highlighted in red color)*

Listing 1: preprocess in label_image of tensorflow

```
def read_tensor_from_image_file(file_name, input_height=299, input_width=299,
                input_mean=0, input_std=255):
  input_name = "file_reader"
  output_name = "normalized"
  file_reader = tf.read_file(file_name, input_name)
  if file_name.endswith(".png"):
    image_reader = tf.image.decode_png(file_reader, channels = 3,
                                       name='png_reader')
  elif file_name.endswith(".gif"):
    image_reader = tf.squeeze(tf.image.decode_gif(file_reader,
                                                  name='gif_reader'))
  elif file_name.endswith(".bmp"):
    image_reader = tf.image.decode_bmp(file_reader, name='bmp_reader')
  else:
    image_reader = tf.image.decode_jpeg(file_reader, channels = 3,
                                        name='jpeg_reader')
  float_caster = tf.cast(image_reader, tf.float32)
  dims_expander = tf.expand_dims(float_caster, 0);
  resized = tf.image.resize_bilinear(dims_expander, [input_height, input_width])
  normalized = tf.divide(tf.subtract(resized, [input_mean]), [input_std])
  sess = tf.Session()
  result = sess.run(normalized)

  return result
```

Listing 2: preprocess in cppclassification of Caffe

```
void Classifier::Preprocess(const cv::Mat& img,
                            std::vector<cv::Mat>* input_channels) {
  /* Convert the input image to the input image format of the network. */
  cv::Mat sample;
  if (img.channels() == 3 && num_channels_ == 1)
    cv::cvtColor(img, sample, cv::COLOR_BGR2GRAY);
  else if (img.channels() == 4 && num_channels_ == 1)
    cv::cvtColor(img, sample, cv::COLOR_BGRA2GRAY);
  else if (img.channels() == 4 && num_channels_ == 3)
    cv::cvtColor(img, sample, cv::COLOR_BGRA2BGR);
  else if (img.channels() == 1 && num_channels_ == 3)
    cv::cvtColor(img, sample, cv::COLOR_GRAY2BGR);
  else
    sample = img;

  cv::Mat sample_resized;
  if (sample.size() != input_geometry_)
    cv::resize(sample, sample_resized, input_geometry_);
  else
    sample_resized = sample;

  cv::Mat sample_float;
  if (num_channels_ == 3)
    sample_resized.convertTo(sample_float, CV_32FC3);
  else
    sample_resized.convertTo(sample_float, CV_32FC1);

  cv::Mat sample_normalized;
  cv::subtract(sample_float, mean_, sample_normalized);

  /* This operation will write the separate BGR planes directly to the
   * input layer of the network because it is wrapped by the cv::Mat
   * objects in input_channels. */
  cv::split(sample_normalized, *input_channels);

  CHECK(reinterpret_cast<float*>(input_channels->at(0).data)
        == net_->input_blobs()[0]->cpu_data())
    << "Input channels are not wrapping the input layer of the network.";
}
```

Listing 3: Imagenet classification with Torch7

```
function preprocess(im, img_mean)
  -- rescale the image
  local im3 = image.scale(im,224,224,'bilinear')
  -- subtract imagenet mean and divide by std
  for i=1,3 do im3[i]:add(-img_mean.mean[i]):div(img_mean.std[i]) end
  return im3
end
```

Listing 4: Imagenet classification with PyTorch

```
def main():
    global args, best_prec1
    args = parser.parse_args()
...
    # Data loading code
    traindir = os.path.join(args.data, 'train')
    valdir = os.path.join(args.data, 'val')
    normalize = transforms.Normalize(mean=[0.485, 0.456, 0.406],
                                     std=[0.229, 0.224, 0.225])

    train_dataset = datasets.ImageFolder(
        traindir,
        transforms.Compose([
            transforms.RandomResizedCrop(224),
            transforms.RandomHorizontalFlip(),
            transforms.ToTensor(),
            normalize,
        ]))

    if args.distributed:
        train_sampler = torch.utils.data.distributed.DistributedSampler(train_dataset)
    else:
        train_sampler = None

    train_loader = torch.utils.data.DataLoader(
        train_dataset, batch_size=args.batch_size, shuffle=(train_sampler is None),
        num_workers=args.workers, pin_memory=True, sampler=train_sampler)

    val_loader = torch.utils.data.DataLoader(
        datasets.ImageFolder(valdir, transforms.Compose([
            transforms.Resize(256),
            transforms.CenterCrop(224),
            transforms.ToTensor(),
            normalize,
        ])),
        batch_size=args.batch_size, shuffle=False,
        num_workers=args.workers, pin_memory=True)
```

Listing 5: code snippnet in deepdetect

```
    int read_file(const std::string &fname)
    {
      cv::Mat img = cv::imread(fname,_bw ? CV_LOAD_IMAGE_GRAYSCALE :
                                           CV_LOAD_IMAGE_COLOR);
      if (img.empty())
    {
      LOG(ERROR) << "empty image";
      return -1;
    }
      _imgs_size.push_back(std::pair<int,int>(img.rows,img.cols));
      cv::Size size(_width,_height);
```

```
    cv::Mat rimg;
    cv::resize(img,rimg,size,0,0,CV_INTER_CUBIC);
    _imgs.push_back(rimg);
    return 0;
  }
```

### B. Software Version and Model Information for Attack Demonstration

Here we present the software setup for the attack demonstration. Although the example used here targets applications from the Caffe platform, the risk is not limited to Caffe. We have tested the scaling functions in Caffe, TensorFlow and Torch. All of them are vulnerable to downscaling attacks.

The Caffe package and the corresponding image classification examples were checked-out directly from the official GitHub on October 25, 2017, and the OpenCV used was the latest stable version from the following URL: https://github.com/opencv/opencv/archive/2.4.13.4.zip

We used the BVLC CaffeNet Model in our proof of concept exploitation. The model is the result of training based on the instructions provided by the instruction in the original Caffe package. To avoid any mistakes in model setup, we download the model file directly from BVLC's official GitHub page. Detail information about the model is provided in the list below.

Listing 6: Image Classifier Model

```
name: BAIR/BVLC GoogleNet Model
caffemodel: bvlc_googlenet.caffemodel
caffemodel_url: http://dl.caffe.berkeleyvision.org/bvlc_googlenet.caffemodel
caffe_commit: bc614d1bd91896e3faceaf40b23b72dab47d44f5
```

### C. Command Lines

The downscaling threat was demonstrated based on the default Caffe example CPPClassification. The exact command line was shown in the list below.

Listing 7: Image Classification Command Line

```
./classification.bin models/bvlc_googlenet/deploy.prototxt
models/bvlc_googlenet/bvlc_googlenet.caffemodel
data/ilsvrc12/imagenet_mean.binaryproto
data/ilsvrc12/synset_words.txt
IMAGE_FILE
```

### D. Sample Output

The lists below is the classification results for the sample images used in the example section.

Listing 8: Sample Classification Results

```
# wolf-in-sheep.png [Image size: 224*672]
./classification.bin  models/bvlc_googlenet/deploy.prototxt
models/bvlc_googlenet/bvlc_googlenet.caffemodel
data/ilsvrc12/imagenet_mean.binaryproto  data/ilsvrc12/synset_words.txt
/tmp/sample/wolf-in-sheep.png
---------- Prediction for /tmp/sample/wolf-in-sheep.png ----------
0.8890 - "n02114548 white wolf, Arctic wolf, Canis lupus tundrarum"
0.0855 - "n02120079 Arctic fox, white fox, Alopex lagopus"
0.0172 - "n02134084 ice bear, polar bear, Ursus Maritimus, Thalarctos maritimus"
0.0047 - "n02114367 timber wolf, grey wolf, gray wolf, Canis lupus"
0.0019 - "n02111889 Samoyed, Samoyede"


# wolf.png [Image size: 224*224]
./classification.bin  models/bvlc_googlenet/deploy.prototxt
models/bvlc_googlenet/bvlc_googlenet.caffemodel
data/ilsvrc12/imagenet_mean.binaryproto  data/ilsvrc12/synset_words.txt
/tmp/sample/wolf.png
---------- Prediction for /tmp/sample/wolf.png ----------
0.8890 - "n02114548 white wolf, Arctic wolf, Canis lupus tundrarum"
0.0855 - "n02120079 Arctic fox, white fox, Alopex lagopus"
0.0172 - "n02134084 ice bear, polar bear, Ursus Maritimus, Thalarctos maritimus"
0.0047 - "n02114367 timber wolf, grey wolf, gray wolf, Canis lupus"
0.0019 - "n02111889 Samoyed, Samoyede"


# cat-in-sheep.png [Image size: 224*672]
./classification.bin  models/bvlc_googlenet/deploy.prototxt
models/bvlc_googlenet/bvlc_googlenet.caffemodel
data/ilsvrc12/imagenet_mean.binaryproto  data/ilsvrc12/synset_words.txt
/tmp/sample/cat-in-sheep.png
---------- Prediction for /tmp/sample/cat-in-sheep.png ----------
0.1312 - "n02127052 lynx, catamount"
0.1103 - "n02441942 weasel"
0.1068 - "n02124075 Egyptian cat"
0.1000 - "n04493381 tub, vat"
0.0409 - "n04209133 shower cap"


# cat.png [Image size: 224*224]
./classification.bin  models/bvlc_googlenet/deploy.prototxt
models/bvlc_googlenet/bvlc_googlenet.caffemodel
data/ilsvrc12/imagenet_mean.binaryproto  data/ilsvrc12/synset_words.txt
/tmp/sample/cat.png
---------- Prediction for /tmp/sample/cat.png ----------
0.1312 - "n02127052 lynx, catamount"
0.1103 - "n02441942 weasel"
0.1068 - "n02124075 Egyptian cat"
0.1000 - "n04493381 tub, vat"
0.0409 - "n04209133 shower cap"
```